\documentclass[12pt,english]{article}

\usepackage{natbib}
\usepackage{url}
\makeatletter
\def\url@leostyle{%
    \def\UrlFont{\sf}}{\def\UrlFont{\small\ttfamily}}
\makeatother
\usepackage{hyperref}

\usepackage{authblk}

\usepackage{graphicx}
\usepackage{babel}
\usepackage{amsmath}
\newcommand{\alg}{\mathcal}
\newcommand{\ket}[1]{|#1\rangle}
\newcommand{\bra}[1]{\langle#1|}


\usepackage{geometry}
\raggedright

\usepackage{times}
\usepackage[T1]{fontenc}
\usepackage[latin1]{inputenc}

\usepackage{setspace}
\usepackage[]{footmisc}

\usepackage[font=normalsize,labelfont=bf]{caption}

\usepackage{titlesec}
\titleformat*{\section}{\bf}
\titleformat*{\subsection}{\it}
\titleformat*{\subsubsection}{\it}

\begin{document}

\title{On the Debate Concerning the Proper
  Characterisation of Quantum Dynamical Evolution}
\author[$\dagger$]{Michael E. Cuffaro}
\author[*]{Wayne C. Myrvold}
\affil[$\dagger$,*]{The University of Western Ontario, Department of
  Philosophy}

\maketitle

\thispagestyle{empty}

\begin{abstract}
\noindent There has been a long-standing and sometimes passionate
debate between physicists over whether a dynamical framework for
quantum systems should incorporate not completely positive (NCP) maps
in addition to completely positive (CP) maps. Despite the
reasonableness of the arguments for complete positivity, we argue that
NCP maps should be allowed, with a qualification: these should be
understood, not as reflecting `not completely positive' evolution, but
as linear extensions, to a system's entire state space, of CP maps that
are only partially defined. Beyond the domain of definition of a
partial-CP map, we argue, much may be permitted.
\end{abstract}


\pagebreak

\section{Introduction}
\label{sec:intro}

Conventional wisdom has it that any evolution of a quantum system can
be represented by a family of completely positive (CP) maps on its
state space. Moreover, there seem to be good arguments that evolutions
outside this class must be regarded as unphysical. But orthodoxy is
not without dissent; several authors have argued for considering
evolutions represented by maps that are not completely positive
(NCP).

The debate has implications that have the potential to go deep. The
possibility of incorporating NCP maps into our quantum dynamical
framework may illuminate much regarding the nature of and relation
between quantum entanglement and other types of quantum correlations
\citep[]{devi2011}. If the use of NCP maps is illegitimate however,
such investigations must be dismissed without further ado.

In the following, we will argue for the proposition that NCP maps
should be allowed\textemdash but we will add a caveat: one should not
regard NCP dynamical maps as descriptions of the `not completely
positive evolution' of quantum systems. An `NCP map', properly
understood, is a linear extension, to a system's entire state space,
of a CP map that is only defined on a subset of this state space.
In fact, as we will see, not much constrains the extension of a
partially defined CP map. Depending on the characteristics
of the state preparation, such extensions may be not completely
positive, inconsistent,\footnote{Strictly speaking, when an
  inconsistent map is used this should not be seen as an extension but
  as a change of state space. This will be clarified below.} or even
nonlinear.

The paper will proceed as follows: in Section \ref{sec:evol} we review
the essential aspects of the theory of open quantum systems and in
Section \ref{sec:cpmap} we present the standard argument for complete
positivity. In Section \ref{sec:deb} we consider the issues involved
in the debate over NCP maps and in Section \ref{sec:lccp} we present
our interpretation of the debate and what we believe to be its
resolution.

\section{Evolution of a Quantum System}
\label{sec:evol}
Consider a quantum system $S$ that is initially in a state $\rho_S^0$,
represented by a density operator $\hat{\rho}_S^0$.  If the system is
isolated, its evolution will be given by a one-parameter family of
unitary operators $\{U^t\}$, via
\begin{equation}
\hat{\rho}_S^t = U^t \: \hat{\rho}_S^0 \: U^{\dag t}.
\end{equation}
Suppose, now, that the system interacts with another system $R$, which
may include some piece of experimental apparatus.  We take $R$ to
include everything with which $S$ interacts.  Suppose that $S$ is
prepared in a state that is uncorrelated with the state of $R$ (though
it may be entangled with some other system, with which it doesn't
interact), so that the initial state of the composite system $S+R$ is
\begin{equation}
\hat{\rho}_{SR}^0 = \hat{\rho}_S^0 \otimes \hat{\rho}_R^0.
\end{equation}
The composite system will evolve unitarily:
\begin{equation}
\hat{\rho}_{SR}^t = U^t \: \hat{\rho}_{SR}^0 \: U^{\dag t},
\end{equation}
where now $\{U^t\}$ is a family of operators operating on the Hilbert
space $\mathcal{H}_S \otimes \mathcal{H}_R$ of the composite
system. It is easy to show (see, e.g., \citealt[\textsection
  8.2.3]{nielsenChuang2000}) that, for each $t$, there will be a set
$\{W_i(t) \}$ of operators, which depend on the evolution operators
$\{ U^t \}$ and the initial state of $R$, such that
\begin{equation}\label{Kraus}
\begin{array}{l}
\hat{\rho}_S^t = \sum_i W_i(t) \: \hat{\rho}_S^0 \: W_i^\dag (t);
\\ \\
\sum_i W_i^\dag (t)  W_i(t) = I.
\end{array}
\end{equation}
This is all in the Schr\"odinger picture, in which we represent a
change of state by a change in the density operator used.  We can also
use the Heisenberg picture, which represents a state change via a
transformation of the algebra of operators used to represent
observables:
\begin{equation}
\rho_{S}^t(A) = \rho_S^0(A^t),
\end{equation}
where
\begin{equation}
A^t = \sum_i W_i(t) A^0 W_i^\dag(t).
\end{equation}

In addition to unitary evolution of an undisturbed system, we also
associate state changes with measurements, via the collapse postulate.
In the case of a von Neumann measurement, there is a complete set
$\{P_i \}$ of projections onto the eigenspaces of the observable
measured, and the state undergoes one of the state transitions $T_i$
given by
\begin{equation}
\mathcal{T}_i \hat{\rho} = \frac{P_i \: \hat{\rho} \:Pi}{\mbox{Tr}(P_i
  \:\hat{\rho})},
\end{equation}
The probability that the state transition will be $\mathcal{T}_i$ is
${\mbox{Tr}(P_i \:\hat{\rho})}$.
When a measurement has been performed, and we don't yet know the
result, the state that represents our state of knowledge of the system
is
\begin{equation}
\mathcal{T} \hat{\rho} = \sum_i P_i \: \hat{\rho} \: P_i.
\end{equation}
Note that this, also, has the form (\ref{Kraus}).

One can also consider \emph{selective} operations, that is, operations
that take as input a state and yield a transformed state, not with
certainty, but with some probability less than one, and fail,
otherwise. One such operation is the procedure of performing a
measurement and keeping the result only if the outcome lies in a
specified set (for example, we could do a spin measurement and select
only `+' outcomes); the operation fails (does not count as preparing a
state at all) if the measurement yields some other result. A selective
operation is represented by a transformation of  the state space that
does not preserve norm. A selective operation $\mathcal{T}$, applied
to state $\rho$, produces a final state $\mathcal{T}\rho$ with
probability $\mathcal{T}\rho(I)$, and no result otherwise.

Unitary evolution, evolution of a system interacting with an
environment with which it is initially correlated, and
measurement-induced collapse can all be represented in the form
(\ref{Kraus}).  The class of state transformations that can be
represented in this form is precisely the class of \emph{completely
  positive} transformations of the system's state space, to be
discussed in the next section.

\section{Completely Positive Maps}
\label{sec:cpmap}
We will want to consider, not just transformations of a single
system's state space, but also mappings from one state space to
another.  The operation of forming a reduced state by tracing out the
degrees of freedom of a subsystem is one such mapping; as we will see
below, assignment maps used in the theory of open systems are
another.

We associate with any quantum system a $C^*$-algebra whose
self-adjoint elements represent the observables of the system.  For
any $C^*$-algebra $\alg{A}$, let $\alg{A}^*$ be its dual space, that
is, the set of bounded linear functionals on $\alg{A}$.  The state
space of $\alg{A}$, $\mathcal{K}(\alg{A})$, is the subset of
$\alg{A}^*$  consisting of positive linear functionals of unit norm.

For any linear mapping  $\mathcal{T}: \alg{A} \rightarrow \alg{B}$,
there is a dual map $\mathcal{T^*}: \alg{A}^* \rightarrow \alg{B}^*$,
defined by
\begin{equation}
\mathcal{T}^*  \mu(A) = \rho(\mathcal{T} A) \mbox{ for all } A \in \alg{A}.
\end{equation}
If $\mathcal{T}$ is positive and unital, then $\mathcal{T}^*$  maps
states on $\alg{A}$ to states on $\alg{B}$.  Similarly, for any
mapping of the state space of one algebra into the state space of
another, there is a corresponding dual map on the algebras.

For any $n$, let $W_n$ be an $n$-state system that doesn't interact
with our system $S$, though it may be entangled with $S$. Given a
transformation $\mathcal{T}$ of the state space of $S$, with
associated transformation $\mathcal{T}$ of $S$'s algebra, we can
extend this transformation to one on the state space of the composite
system $S + W_n$,  by stipulating that the transformation act
trivially on observables of $W_n$.
\begin{equation} 
(\mathcal{T}^* \otimes I_n) \rho(A \otimes B) = \rho(\mathcal{T}(A)
  \otimes B).
\end{equation}
A mapping $\mathcal{T}^*$ is \emph{$n$-positive }if $\mathcal{T}^*
\otimes I_n$ is positive, and \emph{completely positive} if it is
$n$-positive for all $n$.  If $S$ is a $k$-state system, a
transformation of $S$'s state space is completely positive if it is
$k$-positive.

It can be shown \citep[\textsection 8.2.4]{nielsenChuang2000} that,
for any completely positive map $\mathcal{T}^*: \mathcal{K}(\alg{A})
\rightarrow \mathcal{K}(\mathcal{B})$, there are operators $W_i :
\mathcal{H}_\alg{A} \rightarrow \mathcal{H}_\alg{B}$ such that
\begin{equation}
\begin{array}{l}
\mathcal{T}^* \rho (A) = \rho(\sum_i W_i^\dag \: A \: W_i);
\\
\\
\sum_i W_i^\dag W_i \leq I.
\end{array}
\end{equation}
This is equivalent to a transformation of density operators
representing the states,
\begin{equation}
\hat{\rho} \rightarrow \hat{\rho}' = \sum_i W_i \: \hat{\rho} \: W_i^\dag.
\end{equation}
The standard argument that any physically realisable operation on
the state of a system $S$  must be completely positive goes as
follows. We should be able to apply the operation $\mathcal{T}^*$ to
$S$ regardless of its initial state, and the effect on the state of
$S$ will be the same whether or not $S$ is entangled with a
``witness'' system $W_n$.  Since $S$ does not interact with the
witness, applying operation $T^*$ to $S$ is equivalent to applying
$\mathcal{T}^* \otimes I_n$  to the composite system $S + W_n$.  Thus,
we require each mapping $\mathcal{T}^* \otimes I_n$ to be a positive
mapping, and this is equivalent to the requirement that
$\mathcal{T}^*$ be completely positive.

To see what goes wrong if the transformation applied to $S$ is
positive but not completely positive, consider the simplest case, in
which $S$ is a qubit.  Suppose that we could apply a transformation
$\rho_S^0 \rightarrow \rho_S^1$ that left the expectation values of
$\sigma_x$ and $\sigma_y$ unchanged, while flipping the sign of the
expectation value of $\sigma_z$.
\begin{equation}\label{flip}
\rho_S^1( \sigma_x) = \rho_S^0(\sigma_x); \quad  \rho_S^1( \sigma_y) =
\rho_S^0(\sigma_y);  \quad \rho_S^1( \sigma_z) = -\rho_S^0(\sigma_z).
\end{equation}
Suppose that $S$ is initially entangled with another qubit, in,
e.g., the singlet state, so that
\begin{equation}
\rho_{SW}^0(\sigma_x \otimes \sigma_x) = \rho_{SW}^0(\sigma_y \otimes
\sigma_y) = \rho_{SW}^0(\sigma_z \otimes \sigma_z) = -1.
\end{equation}
If we could apply the  transformation (\ref{flip})  to $S$ when it is
initially in a singlet state with $W$,  this would result in a state
$\rho_{SW}^1$ of $S + W$ satisfying,
\begin{equation}\label{badstate}
\rho_{SW}^1(\sigma_x \otimes \sigma_x) = \rho_{SW}^1(\sigma_y \otimes
\sigma_y) =  -1; \quad \rho_{SW}^1(\sigma_z \otimes \sigma_z) = +1.
\end{equation}
This is disastrous.  Suppose we do a Bell-state measurement.  One of
the possible outcomes is the state $\ket{\Psi^+}$, and the projection
onto this state is
\begin{equation}
\ket{\Psi^+} \bra{\Psi^+} = \frac{1}{4} \left(I + \sigma_x \otimes
\sigma_x +  \sigma_y \otimes \sigma_y -  \sigma_z \otimes \sigma_z
\right).
\end{equation}
A state satisfying (\ref{badstate}) would assign an expectation value
of $-1/2$ to this projection operator, rendering it impossible to
interpret this expectation value as the probability of a Bell-state
measurement resulting in $\ket{\Psi^+}$.

Note that the set-up envisaged in the argument is one in which it is
presumed that we can prepare the system $S$ in a state that is
uncorrelated with the active part of its environment $R$.  This set-up
includes the typical laboratory set-up, in which system and apparatus
are prepared independently in initial states; it also includes
situations in which we prepare a system in an initial state and then
put it into interaction with an environment, such as a heat bath, that
has been prepared independently.

\section{The Debate Concerning Not Completely Positive Dynamical Maps}
\label{sec:deb}

The early pioneering work of \citet[]{sudarshan1961}, and
\citet[]{jordan1961}, did not assume complete positivity, but instead
characterised the most general dynamical framework for quantum systems
in terms of linear maps of density matrices. After the important work
of, for instance, \citet[]{choi1972} and \citet[]{kraus1983}, however,
it became increasingly generally accepted that complete positivity
should be imposed as an additional requirement. Yet despite the
reasonableness of the arguments for complete positivity, the
imposition of this additional requirement was not universally
accepted. Indeed, the issue of whether the more general or the more
restricted framework should be employed remains controversial among
physicists. At times, the debate has been quite passionate
\citep[e.g.,][]{simmons1981,raggio1982,simmons1982}.

The issues involved in the debate were substantially clarified by an
exchange between Pechukas and Alicki which appeared in a series of
papers between 1994 and 1995. Pechukas and Alicki analysed the
dynamical map, $\Lambda$, for a system into three separate components:
an `assignment map', a unitary on the combined state space, and a
trace over the environment:
\begin{align}
\label{eqn:dynmap}
\rho_S \to \Lambda\rho_S = \mbox{tr}_R(U\Phi\rho_SU^\dagger),
\end{align}
with $S,R$ representing the system of interest and the environment
(the `reservoir') respectively, and the assignment map, $\Phi$, given
by
\begin{align}
\label{eqn:assgmap}
\rho_S \to \Phi\rho_S = \rho_{SR}.
\end{align}

Since the unitary and the partial trace map are both CP, whether or
not $\Lambda$ itself is CP is solely determined by the properties of
$\Phi$, the assignment map. $\Phi$ represents an assignment of
`initial conditions' to the combined system: it assigns a
\emph{single} state, $\rho_{SR}$, to each state $\rho_S$. My use of
inverted commas here reflects the fact that such a unique assignment
cannot be made in general, since in general the state of the reservoir
will be unknown. It will make sense to use such a map in some cases,
however; for instance if there is a class $\Gamma$ of possible initial
states $S+R$ that is such that, within this class, $\rho_S$ uniquely
determines $\rho_{SR}$. Or it might be that, even though there are
distinct possible initial states in $\Gamma$ that yield the same
reduced state $\rho_S$, the evolution of $\rho_S$ is (at least
approximately) insensitive to which of these initial states is the
actual initial conditions.

When $\Phi$ is linear:
\begin{align}
\Phi(\lambda\rho_1 + (1-\lambda)\rho_2) = \lambda\Phi(\rho_1) +
(1-\lambda)\Phi(\rho_2),
\end{align}
consistent:
\begin{align}
\mbox{tr}_R(\Phi\rho_S) = \rho_S,
\end{align}
and of product form, one can show that $\Phi$ is of necessity CP as
well. \citet[]{pechukas1994} inquired into what follows from the
assumption that $\Phi$ is linear, consistent, and positive. Pechukas
showed that if $\Phi$ is defined everywhere on the state space, and is
linear, consistent, and positive, \emph{it must be a product map}:
$\rho_S \xrightarrow{\Phi} \rho_{SR} = \rho_S \otimes \rho_R$, with
$\rho_R$ a fixed density operator on the state space of the reservoir
(i.e., all $\rho_S$'s are assigned the same $\rho_R$). This is
undesirable as there are situations in which we would like to describe
the open dynamics of systems that do not begin in a product state with
their environment. For instance, consider a multi-partite entangled
state of some number of qubits representing the initial conditions of
a quantum computer, with one of the qubits representing a `register'
and playing the role of $S$, and the rest playing the role of the
reservoir $R$. If we are restricted to maps that are CP on the
system's entire state space then it seems we cannot describe the
evolution of such a system.

Pechukas went on to show that when one allows correlated initial
conditions, $\Lambda$, interpreted as a dynamical map defined on the
entire state space of $S$, may be NCP. In order to avoid the ensuing
negative probabilities, one can define a `compatibility domain' for
this NCP map; i.e., one stipulates that $\Lambda$ is defined only for
the subset of states of $S$ for which $\Lambda\rho_S \geq 0$ (or
equivalently, $\Phi\rho_S \geq 0$). He writes:

\begin{quote}
The operator $\Lambda$ is defined, via reduction from unitary $S+R$
dynamics, only on a subset of all possible $\rho_S$'s. $\Lambda$ may
be extended\textemdash trivially, by linearity\textemdash to the set
of \emph{all} $\rho_S$, but the motions $\rho_S \to \Lambda\rho_S$ so
defined may not be physically realizable ... Forget complete
positivity; $\Lambda$, extended to all $\rho_S$, may not even be
positive \citeyearpar[]{pechukas1994}.
\end{quote}

In his response to Pechukas, \citet[]{alicki1995} conceded that the
only initial conditions appropriate to an assignment map satisfying all
three ``natural'' requirements\textemdash of linearity, consistency,
and complete positivity\textemdash are product initial
conditions. However, he rejected Pechukas's suggestion that in order
to describe the evolution of systems coupled to their environments one
must forego the requirement that $\Lambda$ be CP on $S$'s entire state
space. Alicki calls this the ``fundamental positivity condition.''
Regarding Pechukas's suggestion that one may use an NCP map with a
restricted compatibility domain, Alicki writes:

\begin{quote}
... Pechukas proposed to restrict ourselves to such initial density
matrices for which $\Phi\rho_S \geq 0$. Unfortunately, it is
impossible to specify such a domain of positivity for a general case,
and moreover there exists no physical motivation in terms of
operational prescription which would lead to [an NCP assignment of
  initial conditions] \citep[]{alicki1995}.
\end{quote}

It is not clear exactly what is meant by Alicki's assertion that it is
impossible to \emph{specify} the domain of positivity of such a map in
general, for does not the condition $\Phi\rho_S \geq 0$ itself
constitute a specification of this domain? Most plausibly, what Alicki
intends is that \emph{determining} the compatibility domain will be
exceedingly difficult for the general case. We will return to this
question in the next section, as well as to the question of the
physical motivation for utilising NCP maps.

In any case, rather than abandoning the fundamental positivity
condition, Alicki submits that in situations where the system and
environment are initially correlated one should relax either
consistency or linearity. Alicki attempts to motivate this by arguing
that in certain situations the preparation process may induce an
instantaneous perturbation of $S$. One may then define an inconsistent
or nonlinear, but still completely positive, assignment map in which
this perturbation is represented.

According to \citet[]{pechukas1995}, however, there is an important
sense in which one should not give up the consistency
condition. Consider an inconsistent linear assignment map that takes
the state space of $S$ to a convex subset of the state space of
$S+R$. Via the partial trace it maps back to the state space of $S$,
but since the map is not necessarily consistent, the traced out state,
$\rho_S'$, will not in general be the same as $\rho_S$; i.e.,
\begin{align}
\rho_S \xrightarrow{\Phi} \Phi\rho_S \xrightarrow{\mbox{tr}_R}
\rho_S' \neq \rho_S.
\end{align}

Now each assignment of initial conditions, $\Phi\rho_S$, will
generate a trajectory in the system's state space which we can regard
as a sequence of CP transformations of the form:
\begin{align}
\rho_S(t) = \mbox{tr}_R(U_t\Phi\rho_SU_t^\dagger).
\end{align}
At $t=0$, however, the trajectory begins from $\rho_S'$, not
$\rho_S$. $\rho_S$, in fact, is a fixed point that lies \emph{off} the
trajectory. This may not be completely obvious, prima facie, for is it
not the case, the sceptical reader might object, that we can describe
the system as evolving from $\rho_S$ to $\rho_{SR}$ via the assignment
map and then via the unitary transformation to its final state? While
this much may be true, it is important to remember that $\Phi$ is
supposed to represent an assignment of \emph{initial conditions} to
$S$. On this picture the evolution through time of $\Phi\rho_S$ is a
proxy for the evolution of $\rho_S$. When $\Phi$ is consistent,
$\mbox{tr}_R(U\Phi\rho_SU^\dagger) = \mbox{tr}_R(U\rho_{SR}U^\dagger)$
and there is no issue; however when $\Phi$ is inconsistent,
$\mbox{tr}_R(U\Phi\rho_SU^\dagger) \neq
\mbox{tr}_R(U\rho_{SR}U^\dagger)$, and we can no longer claim to be
describing the evolution of $\rho_S$ through time but only the
evolution of the distinct state $\mbox{tr}(\Phi\rho_S) = \rho_S'$. And
while the evolution described by the dynamical map $\rho'_S(0)
\xrightarrow{\Lambda} \rho'_S(t)$ is completely positive, it has
\emph{not} been shown that the transformation
$\rho_S(0)\xrightarrow{\Lambda} \rho_S(t)$ must always be so.

What of Alicki's suggestion to drop the linearity condition on the
assignment map? It is unclear that this can be successfully physically
motivated, for it is prima facie unclear just what it would mean to
accept nonlinearity as a feature of reduced dynamics. Bluntly put,
quantum mechanics is linear in its standard formulation: the
Schr\"odinger evolution of the quantum-mechanical wave-function is
linear evolution. Commenting on the debate, \citet[]{rodriguez2010}
write: ``giving up linearity is not desirable: it would disrupt
quantum theory in a way that is not experimentally
supported.''

\section{Linearity, Consistency, and Complete Positivity}
\label{sec:lccp}

We saw in the last section that there are good reasons to be sceptical
with respect to the legitimacy of violating any of the three natural
conditions on assignment maps. We will now argue that there are
nevertheless, in many situations, good, physically motivated, reasons
to violate these conditions.

Let us begin with the CP requirement. \emph{Pace} Alicki, one finds a
clear physical motivation for violating complete positivity if one
notes, as \citet[]{shaji2005} do, that if the system $S$ is initially
entangled with $R$, then not all initial states of $S$ are
allowed\textemdash for instance, $\rho_S = \mbox{tr}_R\rho_{SR}$
cannot be a pure state, since the marginal of an entangled state is
always a mixed state. Such states will be mapped to negative matrices
by a linear, consistent, NCP map. On the other hand the map will be
positive for all of the valid states of $S$; this is the so-called
compatibility domain of of the map: the subset of states of $S$ that
are compatible with $\Lambda$.

In light of this we believe it unfortunate that such maps have come to
be referred to as NCP maps, for strictly speaking it is not the map
$\Lambda$ but its linear extension to the entire state space of $S$
that is NCP. $\Lambda$ is indeed CP \emph{within its compatibility
  domain}. In fact this misuse of terminology is in our view at least
partly responsible for the sometimes acrid tone of the debate. From
the fact that the linear extension of a partially defined CP map is
NCP, it does not follow that ``reduced dynamics need not be completely
positive.''\footnote{This is the title of Pechukas's
  \citeyear[]{pechukas1994} article.} Alicki and others are right to
object to this latter proposition, for given the arguments for
complete positivity it is right to demand of a dynamical map that it
be CP on the domain within which it is defined. On the other hand it
is \emph{not} appropriate to insist with Alicki that a dynamical map
must be CP on the entire state space of the system of
interest\textemdash come what may\textemdash for negative
probabilities will only result from states that cannot be the initial
state of the system. Thus we believe that `NCP maps'\textemdash or
more appropriately: \emph{Partial-CP} maps with NCP linear
extensions\textemdash can and should be allowed within a quantum
dynamical framework.

What of Alicki's charge that the compatibility domain is impossible to
``specify'' in general? In fact, the determination of the
compatibility domain is a well-posed problem
\citep[cf.][]{jordan2004}; however, as Alicki alludes to, there may be
situations in which actually determining the compatibility domain will
be computationally exceedingly difficult. But in other
cases\footnote{For examples, see
  \citet[]{jordan2004,shaji2005}.}\textemdash when computing the
compatibility domain \emph{is} feasible\textemdash we see no reason
why one should bar the researcher from using a Partial-CP map whose
linear extension is NCP if it is useful for her to do so. Indeed,
given the clear physical motivation for it, this seems like the most
sensible thing to do in these situations.

There may, on the other hand, be other situations where proceeding in
this way will be inappropriate. For instance, consider a correlated
bipartite system $S+R$ with the following possible initial states:
\begin{align}
\label{eqn:posin}
x_+ \otimes \psi_+,\quad x_- \otimes \psi_-,\quad z_+ \otimes
\phi_+,\quad z_- \otimes \phi_-.
\end{align}
The domain of definition of $\Phi$ consists of the four states
$\{x_+,x_-,z_+,z_-\}$. Suppose we want to extend $\Phi$ so that it is
defined on all mixtures of these states, and is linear. The totally
mixed state of $S$ can be written as an equally weighted mixture of
$x_+$ and $x_-$, and also as an equally weighted mixture of $z_+$ and
$z_-$.
\begin{align}
\label{eqn:mix}
\frac{1}{2}I = \frac{1}{2}x_+ + \frac{1}{2}x_- = \frac{1}{2}z_+ +
\frac{1}{2}z_-.
\end{align}
If $\Phi$ is defined on this state, and is required to be a linear
function, we must have
\begin{align}
\label{eqn:lin}
\Phi(\frac{1}{2}I) & = \frac{1}{2}\Phi(x_+) + \frac{1}{2}\Phi(x_-)
\nonumber \\
& = \frac{1}{2}x_+ \otimes \psi_+ + \frac{1}{2}x_- \otimes \psi_-, \\
\nonumber \\
\Phi(\frac{1}{2}I) & = \frac{1}{2}\Phi(z_+) + \frac{1}{2}\Phi(z_-)
\nonumber \\
& = \frac{1}{2}z_+ \otimes \phi_+ + \frac{1}{2}z_- \otimes \phi_-,
\end{align}
from which it follows that
\begin{align}
\label{eqn:follows}
\frac{1}{2}x_+ \otimes \psi_+ + \frac{1}{2}x_- \otimes \psi_- =
\frac{1}{2}z_+ \otimes \phi_+ + \frac{1}{2}z_- \otimes \phi_-,
\end{align}
which in turn entails that
\begin{align}
\label{eqn:entails}
\psi_+ = \psi_- = \phi_+ = \phi_-,
\end{align}
so $\Phi$ cannot be extended to a linear map on the entire state space
of $S$ unless it is a product map.

It would be misleading to say that assignment maps such as these
violate linearity, for much the same reason as it would be misleading
to say that Partial-CP maps with NCP linear extensions violate
complete positivity. It is not that these maps are defined on a convex
domain, and are nonlinear on that domain; rather, there are mixtures
of elements of the domain on which the function is undefined. But
since we cannot be said to have violated linearity, then \emph{pace}
\citeauthor[]{rodriguez2010}, in such situations we see no reason to
bar the researcher from utilising these `nonlinear' maps, for properly
understood, they are partial-linear maps with nonlinear extensions.

\emph{Pace} Pechukas, there may even be situations in which it is
appropriate to use an inconsistent assignment map. Unlike the previous
cases, in this case the assignment map will be defined on the system's
entire state space. This will have the disadvantage, of course, that our
description of the subsequent evolution will not be a description of the
true evolution of the system, but in many situations one can imagine
that the description will be ``close enough,'' i.e., that
\begin{align}
\mbox{tr}_R(U_t\rho_{SR}U_t^\dagger) \approx
\mbox{tr}_R(U_t\rho_{SR}'U_t^\dagger).
\end{align}

\section{Conclusion}
\label{sec:con}

Bohr warned us long ago against extending our concepts, however
fundamental, beyond their domain of applicability. The case we have
just looked at is an illustration of this important point. The debate
over the properties one should ascribe to the extension of a
partially-defined description is a debate over the properties one
should ascribe to a phantom.

Whether or not we must use a map whose extension is nonlinear, or a
map whose linear extension is NCP, or an inconsistent map, is not a
decision that can be made a priori or that can be shown to follow from
fundamental physical principles. The decision will depend on the
particular situation and on the particular state preparation we are
dealing with.

\pagebreak

\bibliographystyle{ChicagoReedWeb}
\bibliography{Bibliography}{}

\end{document}